**Chi-square-based scoring function for categorization of MEDLINE citations**

**A. Kastrin[1], B. Peterlin[1], D. Hristovski[2]**

[1]Institute of Medical Genetics, University Medical Centre Ljubljana, Ljubljana, Slovenia

[2]Institute for Biostatistics and Medical Informatics, Faculty of Medicine, University of Ljubljana, Ljubljana, Slovenia

**Corresponding author**: Andrej Kastrin, Institute of Medical Genetics, University Medical Centre Ljubljana, Slajmerjeva 3, SI-1000 Ljubljana, Slovenia, email: andrej.kastrin@guest.arnes.si

**Principal corresponding author:** Assist. Prof. Dimitar Hristovski, PhD, Institute for Biostatistics and Medical Informatics, Faculty of Medicine, University of Ljubljana, Vrazov trg 2, SI-1104 Ljubljana, Slovenia, e-mail: dimitar.hristovski@mf.uni-lj.si

# Summary

**Objectives:** Text categorization has been used in biomedical informatics for identifying documents containing relevant topics of interest. We developed a simple method that uses a chi-square-based scoring function to determine the likelihood of MEDLINE® citations containing genetic relevant topic.

**Methods:** Our procedure requires construction of a genetic and a nongenetic domain document corpus. We used MeSH® descriptors assigned to MEDLINE citations for this categorization task. We compared frequencies of MeSH descriptors between two corpora applying chi-square test. A MeSH descriptor was considered to be a positive indicator if its relative observed frequency in the genetic domain corpus was greater than its relative observed frequency in the nongenetic domain corpus. The output of the proposed method is a list of scores for all the citations, with the highest score given to those citations containing MeSH descriptors typical for the genetic domain.

**Results:** Validation was done on a set of 734 manually annotated MEDLINE citations. It achieved predictive accuracy of 0.87 with 0.69 recall and 0.64 precision. We evaluated the method by comparing it to three machine learning algorithms (support vector machines, decision trees, naïve Bayes). Although the differences were not statistically significantly different, results showed that our chi-square scoring performs as good as compared machine learning algorithms.

**Conclusions:** We suggest that the chi-square scoring is an effective solution to help categorize MEDLINE citations. The algorithm is implemented in the BITOLA literature-based discovery support system as a preprocessor for gene symbol disambiguation process.

# Introduction

The proliferation of the biomedical literature makes it difficult even for experts to absorb all the relevant knowledge in their specific field of interest. Effective heuristics for identifying articles containing relevant topics would be beneficial both for the individual researchers and for curation of biomedical databases. Manual literature curation is a resource and time-consuming process that is prone to inconsistencies [1]. Therefore, an automated system which could correctly determine the relevant topic of the citation retrieved from a bibliographic database, is needed. This is referred to as text or document categorization (DC), which is a process of assigning a text document to one or more categories based on its content or topic [2].

A large body of studies has been published addressing the problem of DC in biomedical domain. Most frequently used approaches include support vector machines (SVM), decision trees (DT), and naïve Bayes (NB). Humphrey et al. [3] presented the technique of automatic indexing of documents by discipline with broad descriptors that express the general nature and orientation of the document. Donaldson et al. [4] used an SVM algorithm to locate PubMed® citations containing information on protein-protein interaction before they were curated into the Biomolecular Interaction Network Database. Dobrokhotov et al. [5] applied a combination of natural language processing and probabilistic classification to re-rank documents returned by PubMed according to their relevance to Swiss-Prot database curation. Bernhardt et al. [6] developed an automated method for identifying prominent subdomains in medicine that relies on Journal Descriptor Indexing, an automated method for topical categorization of biomedical text. Miotto et al. [7] tested the performance of DT and artificial neural networks to identify PubMed abstracts that contain allergen cross-reactivity information. Chen et al. [8] combined an SVM and a phrase-based clustering algorithm to categorize papers about *Caenorhabditis elegans*. McDonald et al. [9] exploited the maximum entropy classification principle to calculate the likelihood of MEDLINE® abstracts containing quotations of genomic variation data suitable for annotation in mutation databases. Recently, Wang et al. [10] used an NB classifier to speed up the abstract selection process of the Immune Epitope Database reference curation. The most popular platforms to evaluate DC algorithms in biomedical domain are Text Retrieval Conference (TREC) [11] and Critical Assessment of Information Extraction systems in Biology (BioCreAtIvE) [12]. Classical statistical methods and machine learning algorithms have been shown highly useful for DC in the area of biomedical informatics. However, they usually require long computational times and tedious manual preparation of training datasets. Here, we fill this gap by proposing a novel and simple DC algorithm as well as preparation of a training corpus for saving manual annotation.

## Background

The methodology we present here was strongly motivated and is used in support of our previous work on literature-based discovery. BITOLA, a biomedical discovery support system, was designed to discover potentially new relationships between diseases and genes [13, 14]. Gene symbols are short acronyms that often create ambiguities if used outside the context of gene names [15]. For example, in the sentence [16] 'The inverse association between MR and VEGFR-2 expression in carcinoma suggest a potential tumor-suppressive function for MR' we need to decide if 'MR' stands for 'mineralocorticoid receptor' gene or 'magnetic resonance' imaging.

Our method is designed to categorize citations very broadly according to discipline in order to help disambiguate gene symbols in the millions of documents in MEDLINE. It therefore has the potential to filter out nonrelevant (i.e., nongenetics) citations early on. The problem can be formally described as a task of assigning a MEDLINE citation to the genetic or nongenetic domain, based on its content. More specifically, our task was to categorize MEDLINE citations into genetic and nongenetic domain first and then to detect gene symbols only in the genetic domain citations. Here we address only the problem of categorization of MEDLINE citations. We refer to the genetic domain as a subset of MEDLINE citations in which occurrences of gene symbols are more probable than in any other subset of citations.

Therefore, a DC algorithm that has a high categorization speed as well as high classification accuracy is required. Our objective was to investigate the benefit of using the MeSH® controlled vocabulary as a representation level of the MEDLINE citation for DC. We developed and evaluated a document ranking technique based on chi-square test for independence. The chi-square test for DC was introduced first by Oakes et al. [17] and Alexandrov et al. [18]. Preliminary results of our work were presented at AMIA 2008 Annual Symposium [19]. In this work, we highly extend the conference paper with a much more rigorous statistical validation methodology on a larger data set. Our proposed approach is simple to implement and can be easily integrated into the existing framework of the BITOLA system. It is able to process the full MEDLINE distribution in a few hours.

# Methods

## Pre-processing of the corpora

MEDLINE is the main and largest literature database in the biomedical related fields. As of this writing, MEDLINE contains about 17 million research abstracts dated back to the 1950s. MEDLINE citations are manually annotated using MeSH vocabulary by trained indexers from the National Library of Medicine (NLM). MeSH is a controlled vocabulary thesaurus consisting of medical terms at various levels of specificity. There are three types of MeSH terms: main headings (descriptors), supplementary concepts, and qualifiers. Descriptors are the main elements of the vocabulary and indicate the main contents of the article. Qualifiers are assigned to descriptors inside the MeSH fields to express a special aspect of the concept. Each MEDLINE citation is manually assigned around 12 MeSH descriptors. The 2008 MeSH, which was used in this study, contains 24,767 descriptors.

Our statistical procedure requires a genetic and a nongenetic domain corpus. In order to obtain this, we processed the full MEDLINE Baseline Repository, up to the end of 2007, which contains 16,880,015 citations. As the distribution is in XML format, we extracted the relevant elements and transformed them into a relational text format (i.e., one line for each MeSH descriptor occurrence in each citation). A frequency count file was compiled to provide a frequency distribution of MeSH descriptors in the whole MEDLINE corpus.

A subset of articles, which we call a genetic domain corpus, was extracted from the MEDLINE corpus to represent genetically relevant citations. To accomplish this, the 'gene2pubmed' file from the Entrez Gene repository [20] was downloaded and used as a reference list for identifying MEDLINE citations in which gene symbols occur. A frequency count file was then created to provide a frequency distribution of MeSH descriptors in the genetic domain corpus. The citations mentioned in the 'gene2pubmed' file were removed from the MEDLINE frequency distribution in order to provide nongenetic domain corpus.

**Categorization algorithm**

For each of the $k$ MeSH descriptors in the two frequency lists we applied the Pearson's chi-square ($X^2$) test for independence [21] to obtain a statistic, which indicates whether a particular MeSH descriptor $m$ is independent regarding genetic ($G = g$) and nongenetic ($G \neq g$) domain corpus. The $X^2$ test compares the difference between the observed frequencies (i.e., the actual frequencies extracted from corpora) and the expected frequencies (i.e., the frequencies that one would expect by chance alone). The larger the value of $X^2$, the more evidence exists against independence [22].

The complete frequency information needed for the implementation of the $X^2$ test is summarized in Table I and Table II (note that we use the index $i$ to indicate the row of the table and $j$ to indicate the column of the table). Given two corpora $G = g$ and $G \neq g$ we created a $2 \times 2$ contingency table of observed frequencies $O_{ij}$ for each target MeSH descriptor ($M = m$) and other MeSH descriptors ($M \neq m$), as demonstrated in Table I. Here the $O_{11}$ is the frequency of citations in the genetic corpus that the target MeSH descriptor is assigned to and the $O_{21}$ is the frequency of citations in the genetic corpus where the target MeSH descriptor is absent. Likewise, the $O_{12}$ is the frequency of citations in the nongenetic corpus that the target descriptor is assigned to and the $O_{22}$ is the frequency of citations in the nongenetic corpus where the target descriptor is absent. The grand total $N$ is the total of all four frequencies (i.e., $O_{11} + O_{21} + O_{12} + O_{22}$). The row and column totals are denoted with $R$'s and $C$'s with subscripts corresponding to the rows and columns.

-------------------------------------------

Insert Table I about here

-------------------------------------------

Next we calculated the corresponding expected frequencies $E_{ij}$ for each table cell, as demonstrated in Table II.

-------------------------------------------

Insert Table II about here

-------------------------------------------

Given the observed and expected frequencies for each MeSH descriptor in both corpora, the $X^2$ statistic was calculated as defined below [21]:

$$X^2 = \sum_{i=1}^{2} \sum_{j=1}^{2} \frac{\left(O_{ij} - E_{ij}\right)^2}{E_{ij}}$$

If an expected value was less then five, we applied Yates's correction for continuity by subtracting 0.5 from difference between each observed frequency and its expected frequency [22]. The limiting distribution of $X^2$ statistic for a 2 × 2 contingency table is a $\chi^2$ distribution with one degree-of-freedom ($df = 1$). If the $X^2$ is greater than the critical value of 3.84 ($p \leq 0.05$), we can be 95% confident that the particular MeSH descriptor is not independent of the domain and therefore it is more likely a discriminative feature for categorization.

To address the question of the direction of the association between particular MeSH descriptor and domain, we also calculated the indicator variable ($I$) for each descriptor from the same table. A similar approach has been introduced by Oakes et al. [17]. A MeSH descriptor was considered to be a positive indicator (+) if its relative observed frequency in the genetic domain corpus was greater than its relative observed frequency in the nongenetic domain corpus. On the other hand, a MeSH descriptor was considered to be a negative indicator (−) if its relative observed frequency in the genetic domain was lower than its relative observed frequency in the nongenetic domain corpus.

Descriptors that appear highly frequently (e.g., Humans, Animals, Mice, etc.) and are thus not meaningful to the algorithm were removed. We built the list of noninformative MeSH descriptors based on MEDLINE check tags [23].

The categorization algorithm requires two inputs: (i) frequency profiles of all the MeSH descriptors with statistically significant chi-square scores ($X^2 > 3.84$; $p \leq 0.05$), noting which descriptors are positive indicators and which are negative, and (ii) a set of citations to be categorized. The algorithm proceeds by reading each MEDLINE citation in turn and assigning a score to it as follows:

Score = 0

**For each** MeSH descriptor

    **If** MeSH descriptor is a positive indicator

      Score = Score + 1

    **Else if** MeSH descriptor is a negative indicator

      Score = Score − 1

The output of this process is a list of scores for all the citations, with the highest total given to those citations containing MeSH descriptors typical for the genetic domain.

**Benchmark algorithms**

To provide a basis for comparison with the chi-square-based scoring function approach described above, we used three machine learning techniques, including SVM, DT, and NB. Decision to use SVM, DT, and NB classifiers was totally arbitrary, based on a list of top 10 algorithms in machine learning [24]. We refer the reader to [25] for more detailed information about the machine learning algorithms we used. We used SVM implementation in the LIBSVM software library with polynomial kernel [26]. The parameters $\gamma$ and $r$ were set to default value 1. The kernel degree $d$ together with the SVM penalty parameter $C$ were optimized by nested cross-validation over $d$ values {1, 2, 3} and $C$ values {0.01, 1, 100} [27]. For each learning algorithm we conducted four experiments with the following inputs for each MEDLINE citation: (i) title, (ii) abstract, (iii) title and abstract, and (iv) MeSH terms.

The first step is to transform text data into a representation that is suitable for classification methods to use. For title and abstract field citation indexing was performed at the word level by applying a standard vector space model. A vector representation of citations based on word content was created, in which each distinct word is an orthogonal dimension in the vector space (bag-of-words). General English stopwords were removed by using the standard SMART stopword list [28]. After lowercasing the characters, the Lovins stemming algorithm was used to reduce words to their base forms [29]. In the last step we removed all words with document (citation) frequency less than two. Each of the remaining feature stems represents a dimension in the vector space. The feature counts for the citation vectors were then weighted by the TF-IDF scheme [30], which combines the frequency of a term in the citation and in the citations collection as well. The citation vectors were then normalized to unit length, so that abnormally long or short citations did not adversely affect the training process.

**Threshold optimization and performance evaluation**

The manual categorization was considered as a gold standard. In order to create a real world test scenario, we selected a random set of 1,000 MEDLINE articles. Random sampling was performed using NLM's MBR Query Tool [31]. We went through the year 2008 MEDLINE Baseline Distribution and selected documents with Data Completed (DCOM) fields corresponding from 2005/11/16 to 2006/11/14. DCOM field corresponds to the date processing of the MEDLINE citation ends (i.e., when MeSH headings have been added, quality assurance validation has been completed, etc.).

This set of 1,000 citations was manually annotated by two annotators with biological domain knowledge. Their task was to identify citations as being in either the genetic or nongenetic domain. Annotators used the following criteria to determine whether an article is in the genetic domain: (i) the title and/or abstract discuss one or more genes, gene transcripts, gene regulation molecules, DNA, RNA, (ii) genetic diseases and syndromes, (iii) technology, techniques, and methods used for genetic testing. We measured the agreement between the two annotators before the adjudication step using the κ statistic [32]. Consensus voting was then used to achieve complete agreement between judges.

To draw a boundary between genetic and nongenetic domain citations, we plotted predictive accuracy (*Acc*) against score values on a training set, and the threshold parameter (θ) was set to maximize accuracy. Predictive accuracy is the proportion of corrects predictions to the total predictions and is defined as follows:

$$Acc = \frac{TP + TN}{TP + TN + FN + FP},$$

where *TP* is the number of true positive predictions, *TN* the number of true negative predictions, *FP* the number of false positive predictions, and *FN* the number of false negative predictions. The threshold value was estimated by cross-validation. Following the 10-fold cross-validation regime, nine runs were used to optimize threshold and the rest one was used as a test set to evaluate predictive accuracy. Given defined threshold, we then categorized citations in the test set. For example, all citations for which the decision score was greater or equal to the specified threshold value were categorized as genetic domain citations.

Besides accuracy, the performance measures recall (*Rec*), precision (*Pre*), and *F*-measure (*F*) were used to assess the performance of the categorization algorithm:

$$Rec = \frac{TP}{TP + FN},$$

$$Pre = \frac{TP}{TP + FP},$$

$$F = \frac{2 \times Pre \times Rec}{Pre + Rec}.$$

The recall measures the proportion of positive labeled citations (citations are about genetics) that were categorized as positive, and the precision measures the proportion of positive predictions (citations categorized into the genetic domain) that are correct. The $F$-measure is the weighted harmonic mean of precision and recall.

The same cross-validation scheme was used to evaluate the prediction performance of the machine learning algorithms. Values of evaluation measures were averaged over runs for further reporting. McNemar's test [33] was used to test the statistical difference between chi-square-based scoring function algorithm and each of the machine learning algorithm evaluated. This test was performed by summarizing the classification errors of the algorithms and has a low Type I error (the probability of incorrectly detecting a difference when no difference exists).

The computations were carried out in the R software environment for statistical computing and graphics [34].

# Results

The starting point of our algorithm is a frequency profile table of all the MeSH descriptors with statistically significant chi-square scores, noting which descriptors are positive indicators and which are negative. Chi-square feature selection identified many informative MeSH terms, the presence of which suggested a genetic or nongenetic article. 16,891 out of 24,767 MeSH descriptors were statistically significantly different between domains. 3,821 (22.6%) of them were statistical significantly overrepresented in genetic domain corpus, while 13,070 (77.4%) were overrepresented in nongenetic domain corpus. The highest scoring MeSH terms were intuitively reasonable as predictors either of genetics or nongenetic domain citation. For example, the main characteristic terms for genetic domain are 'Molecular Sequence Data', 'Amino Acid Sequence', and 'Base Sequence'.

Our evaluation corpus contained 1,000 citations. Annotators achieved a κ score for inter-annotator agreement of 0.78. The citations that were most difficult to categorize were those concerning general biochemistry, proteomics, and metabolomics. Consensus voting was then used to achieve complete agreement between judges. The annotated corpus is publicly available at our homepage [35] and can serve as a benchmark for other applications. The percentage of genetic citations in this corpus was 18.6% (186 citations about genetics and 814 citations about nongenetic topics). There were 88 citations (8.8%) that contained no MeSH terms and 223 citations without an abstract field (22.3%). The title field was present in all the citations in the evaluation set. We built a join set of citations in which each citation has a title, abstract and MeSH terms. The final set has 734 citations with 173 citations (23.6%) about genetic domain. A total of 10,979 MeSH descriptors were assigned to a set of 734 citations, with average 12.04 MeSH descriptors per citation ($SD$ = 5.15). All further computations were done on that set.

To estimate the quality of the training data and to evaluate the assumption that the citations from the 'gene2pubmed' file were indicative of genetic domain, we joined the citations from the manually annotated evaluation corpus to those derived from the 'gene2pubmed' file. If a citation in the evaluation corpus was present in the 'gene2pubmed' file, we labeled it as genetic citation; otherwise we labeled it as nongenetic citation. The accuracy of 0.77 with 0.98 recall and 0.78 precision was obtained. The $F$-measure was 0.87.

**Results of the categorization algorithm**

The output of the categorization process is a list of decision scores for all the citations. The score distribution for the 734 citations is presented in Figure 1. In Figure 2 the predictive accuracy as a function of score cut-off values is depicted in order to visualize the performance at different points along the decision score distribution. The threshold parameter was set to optimize predictive accuracy, as described earlier in the 'Threshold optimizations' subsection. The maximal training accuracy was 0.85. The corresponding decision score threshold was $\theta = 3$ for all folds. After estimating optimal model threshold on training data, this parameter was used to generate domain predictions of MEDLINE citations.

--------------------------------------------

Insert Figure 1 about here

--------------------------------------------

--------------------------------------------

Insert Figure 2 about here

---------------------------------------------

The performance of the chi-square categorization algorithm was evaluated on the test subset under the cross-validation regime, described earlier in the 'Performance evaluation' subsection. The proposed algorithm achieved a predictive accuracy of 0.87 with 0.69 recall and 0.64 precision. The $F$-measure was 0.66.

**Comparison with machine learning classifiers**

Results on the test data were generated by SVM, DT, and NB classifiers following standard bag-of-words representation of MEDLINE citations using title, abstract, title+abstract or MeSH terms as prediction features. We reduced the feature space from 3,953; 16,722; and 17,095 words to 116; 4,871; and 5,484 unique words for title, abstract, and title+abstract field respectively. The MeSH descriptors vector space of 3,727 features was reduced by applying a list of noninformative MeSH descriptors, resulting in 3,704 unique MeSH descriptors. By this means the noise possibly introduced into the classifier was eliminated, possibly improving its performance.

Table III shows a comparison of the proposed chi-square algorithm with SVM, DT, and NB classifiers. Classification accuracy, recall, precision, and $F$-measure of the 10-fold cross-validations are presented. Results suggest that the proposed chi-square-based algorithm provides more accurate categorization in genetic and nongenetic domain than SVM, DT, and NB. Results were the same when the $F$-measure was used to compare the algorithms.

---------------------------------------------

Insert Table III about here

-------------------------------------------

Table IV displays results for comparing the chi-square scoring function algorithm with the SVM, DT, and NB algorithms for different representations of MEDLINE citations. McNemar's statistics is an average calculated over the 10 runs. Since all $p$-values are greater than 0.05, we cannot reject the null hypothesis which suggest that applying SVM, DT, and NB learning algorithms to construct classifiers for this application can achieve the same classification results. Only in the case of the NB classifier with MeSH descriptors as representative features the chi-square-based scoring function algorithm performs statistically significantly better.

-------------------------------------------

Insert Table IV about here

-------------------------------------------

## Discussion

In this paper, we developed and experimentally validated a semi-supervised DC approach and demonstrated a simple chi-square-based scoring function algorithm. The method has been shown to discover key MeSH descriptors in MEDLINE, which differentiate between genetic and nongenetic domains. The fundamental conclusions of this study can be summarized as follows: (i) in our particular settings the chi-square-based scoring function algorithm proved to be effective for MEDLINE citation categorization task; (ii) the chi-square-based algorithm is as accurate as evaluated machine learning algorithms, namely SVM, DT, and NB, although the differences are not statistically significant; (iii) chi-square-based algorithm is easy to implement into existent text mining systems.

Our chi-square-based algorithm is a simple statistical procedure, based on widely used and well-known chi-square statistics. In spite of the fact that our results did not yield statistical significant differences between the chi-square algorithm and the benchmark machine learning algorithms, they are still promising. Furthermore, the chi-square algorithm does not perform 'transformations' on the original data, possibly affecting reliability of the categorization process. Our approach does not use the full text titles or abstracts, so it is much more efficient. With the exception of the threshold parameter, no other parameters needed to be tuned in the proposed approach. Last but not least, our method does not need a manually tagged training corpus, which is time-consuming, expensive, and error-prone to construct.

The benchmarking experiments indicate that the categorization results are competitive with state-of-the-art machine learning algorithms like SVM, DT, and NB. According to our results, SVM was the second best algorithm, followed by DT. The high dimensional nature of text data has been shown to be the main reason for bad performance of many classification methods [36]. The SVM method is generally the best choice for high dimensional feature representation, such as those for free text, but requires a lot of parameters to adjust and is very time consuming and labor intensive. In addition, the major disadvantage of DT is instability. Small changes in the data often result in a very different tree and big changes in prediction performance. The reason for that is the hierarchical process of tree induction, where errors made in the splits close to the root are carried down the entire tree. However, DT has been overlooked in the domain of text categorization and big advantage of DT over the other classifiers is that they could be visualized as a set of rules explaining the categorization process. NB was the worst performing algorithm, which confirms previous research that NB is a popular but not the best algorithm for DC [37]. However, based on the absence of statistically significant differences between the evaluated approaches we could not be certain that one of the machine learning classifiers used would not have performed better, and MeSH descriptors may not be optimal for the task of identifying genetic citations. However, the results generated with our method are a promising starting point of this task. Following Occam, we doubt that more intricate classifiers should necessarily be preferred over simple approaches. As a heuristic, Occam's razor principle tells us that the simpler model is generally to be preferred over a more sophisticated model [38]. This is particularly relevant in practical online applications where large amounts of unstructured text data have to be processed quickly and accurately.

Our algorithm requires that the input documents to be categorized have assigned MeSH descriptors. That is a limitation for documents without MeSH descriptors. However, our algorithm could be used as a module in a general information extraction or retrieval system by using text words instead of controlled vocabulary indexing terms. Indeed, the immediate goal of this study was to provide a way to categorize biomedical literature according to genetic and nongenetic MEDLINE citations, but the process to categorize text in another domain is fairly straightforward. The researcher must first create a frequency table of words in each of the exploited domains and then run the algorithm over the documents. The initial building process of frequency table is the most time consuming process, but in principle it has to be done only once. Once the indexing table is available, the categorization algorithm is very fast, taking about four hours for all of MEDLINE (based on the dimension scale of the selected MeSH descriptors). We have not evaluated the performance of our algorithm (regarding classification accuracy and process speed) when using text words yet, but we plan do it in the future.

A wide array of statistical and machine learning techniques has been applied to DC. Most of them are based on having some initial set of class-labeled documents, which is used to train an automatic categorization model. There has been much prior work in applying methods to analyzing the biomedical literature to extract biomedical data, but relatively little prior work has addressed the task of screening the entire literature for particular types of citations. To the best of our knowledge no one has tried to generalize the MeSH descriptors in a manner described in this article. Only a few researchers have tested their DC systems using the entire set of MeSH terms. Rubin et al. [39] developed a curation filter, an automated method to identify citations in MEDLINE that contain pharmacogenetics data pertaining to gene-drug relationships. They reported $F$-measures ranging from 0.20 to 0.88 for different experimental settings. DC machinery has also been used in maintaining systematic reviews in the domain of internal medicine [40] and of the efficacy of drug therapy [41].

There are also some disadvantages of our chi-square-based categorization approach. Each MEDLINE citation has an average of twelve MeSH terms assigned. The standard deviation estimated on a collection of 734 randomly selected MEDLINE citations is very high ($SD = 5.15$), suggesting a high scattering of the number of MeSH terms assigned to each citation. Therefore, the citations with more MeSH terms assigned are more prone to extremely positive or extremely negative scores, which means they are categorized either into genetic or nongenetic domain with greater confidence (although the proposed heuristic scoring could not be directly interpreted as a measure of statistical confidence). Many MeSH terms are also too specific to be used as valid representatives of broad topic domains. In addition, the MeSH terms are related hierarchically, and frequently both the parent and the child are assigned to the same citation; this results in artificially increased or decreased chi-square decision score.

## Conclusion and future work

We have proposed a simple chi-square-based scoring method to categorize MEDLINE citations according to its genetic relevance. Results of experimental validation showed that the proposed method is as good as popular machine learning algorithms, although the differences are not statistically significant. Our algorithm could be easily reimplemented as a module in a general information extraction system and may thus be a powerful tool for the broader research community. The algorithm is currently implemented in the BITOLA literature-based discovery support system [42] as a preprocessor for gene symbol disambiguation process.

The presented approach provides very good performance, but further slight modification may allow even better performance. An important aspect that could be addressed in the future is the better selection of representative corpora. The selection of the corpora is a crucial step in the methodologies of corpora linguistics, since it defines the quality of the training dataset. Ideally for our method, the genetic domain corpus should contain information relevant only to genetics. An interesting future research direction is the extension of the proposed methodology for handling simultaneously several genetically specific knowledge sources, which would better reflect the genetic domain. There are several such sources, including Online Mendelian Inheritance in Man (OMIM), Gene Ontology (GO) and Gene Reference Into Function (GeneRIF) databases. Another technique that may increase performance is linking MeSH terms to Unified Medical Language System (UMLS) concepts and semantic types. As future work the setting of the initial weights to characteristic words should also be studied.

## Acknowledgements

We are grateful to Susanne M. Humphrey and Thomas C. Rindflesch of the NLM for helpful suggestions and comments. We would also like to express our sincere gratitude to anonymous reviewers who helped to improve the quality of this paper. This work was supported by the Slovenian Research Agency.

**Table I.** Contingency table of observed frequencies for target MeSH descriptor.

|           | $G = g$   | $G \neq g$ |         |
|-----------|-----------|------------|---------|
| $M = m$   | $O_{11}$  | $O_{12}$   | $= R_1$ |
| $M \neq m$| $O_{21}$  | $O_{22}$   | $= R_2$ |
|           | $= C_1$   | $= C_2$    | $= N$   |

Note: $M$ = MeSH descriptor; $G$ = corpus; $O_{ij}$ = observed frequency; $R_i$ = row total; $C_j$

= column total; $N$ = grand total.

**Table II.** Calculation of expected frequencies for target MeSH descriptor.

|           | $G = g$                      | $G \neq g$                   |
| --------- | ---------------------------- | ---------------------------- |
| $M = m$   | $E_{11} = (R_1 \times C_1) / N$ | $E_{12} = (R_1 \times C_2) / N$ |
| $M \neq m$ | $E_{21} = (R_2 \times C_1) / N$ | $E_{22} = (R_2 \times C_2) / N$ |

Note: $M$ = MeSH descriptor; $G$ = corpus; $E_{ij}$ = expected frequency; $R_i$ = row total of observed frequencies; $C_j$ = column total of observed frequencies; $N$ = grand total of observed frequencies.

**Table III.** Performance of the chi-square-based scoring function algorithm and comparison with SVM, DT, and NB algorithms.

| Algorithm | *Acc* | *Rec* | *Pre* | *F* |
|---|---|---|---|---|
| Chi-Square | 0.87 | 0.69 | 0.64 | 0.66 |
| SVM$_{Title}$ | 0.81 | 0.33 | 0.68 | 0.45 |
| DT$_{Title}$ | 0.79 | 0.20 | 0.74 | 0.31 |
| NB$_{Title}$ | 0.75 | 0.56 | 0.47 | 0.51 |
| SVM$_{Abstract}$ | 0.82 | 0.36 | 0.73 | 0.48 |
| DT$_{Abstract}$ | 0.79 | 0.21 | 0.74 | 0.33 |
| NB$_{Abstract}$ | 0.77 | 0.59 | 0.51 | 0.54 |
| SVM$_{Title+Abstract}$ | 0.81 | 0.35 | 0.72 | 0.48 |
| DT$_{Title+Abstract}$ | 0.78 | 0.16 | 0.65 | 0.26 |
| NB$_{Title+Abstract}$ | 0.76 | 0.59 | 0.49 | 0.53 |
| SVM$_{MeSH}$ | 0.77 | 0.10 | 0.58 | 0.17 |
| DT$_{MeSH}$ | 0.79 | 0.15 | 0.76 | 0.25 |
| NB$_{MeSH}$ | 0.63 | 0.97 | 0.39 | 0.55 |

Note: Chi-Square = proposed chi-square scoring function; SVM = support vector machines; DT = decision trees; NB = naïve Bayes; *Acc* = accuracy; *Rec* = recall; *Pre* = precision; *F* = *F*-measure.

**Table IV.** Results of McNemar's test for comparing the chi-square-based algorithm with the SVM, DT, and NB algorithms.

| | Algorithm | | | | | |
| | Chi-Square − SVM | | Chi-Square − DT | | Chi-Square − NB | |
| Feature | $z_0^2$ | $p$ | $z_0^2$ | $p$ | $z_0^2$ | $p$ |
|---|---|---|---|---|---|---|
| Title | 0.73 | 0.39 | 0.57 | 0.45 | 1.42 | 0.23 |
| Abstract | 0.25 | 0.62 | 0.40 | 0.53 | 1.21 | 0.27 |
| Title + Abstract | 0.44 | 0.51 | 0.80 | 0.37 | 1.33 | 0.25 |
| MeSH | 1.08 | 0.30 | 0.88 | 0.35 | 6.02 | 0.01 |

Note: Chi-Square = proposed chi-square scoring function; SVM = support vector machines; DT = decision trees; NB = naïve Bayes; $z_0^2$ = McNemar's statistic; $p$ = $p$-value of McNemar's statistic.

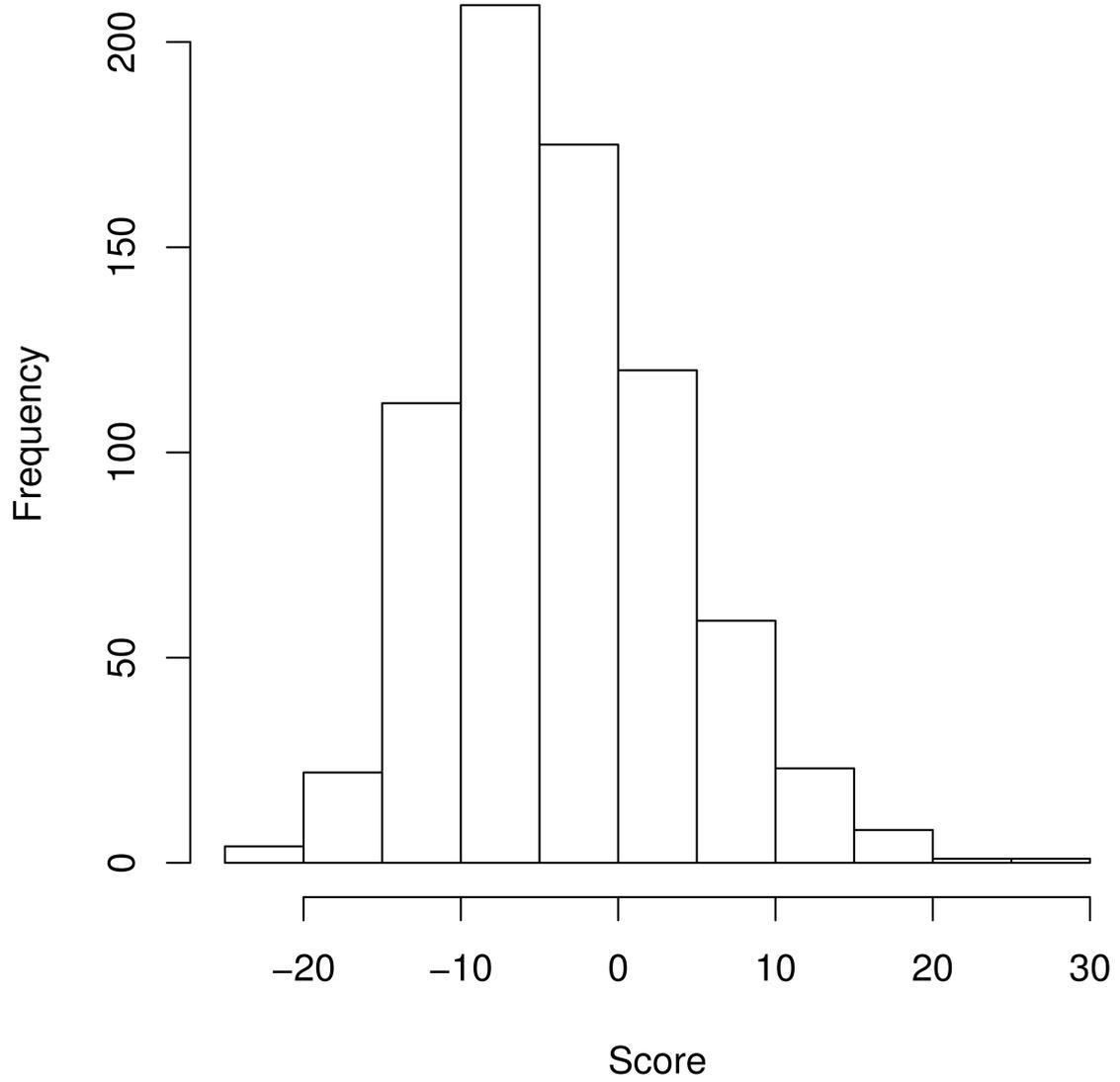

**Figure 1.** Score distribution of the chi-square based scoring function for a set of 734

MEDLINE citations.

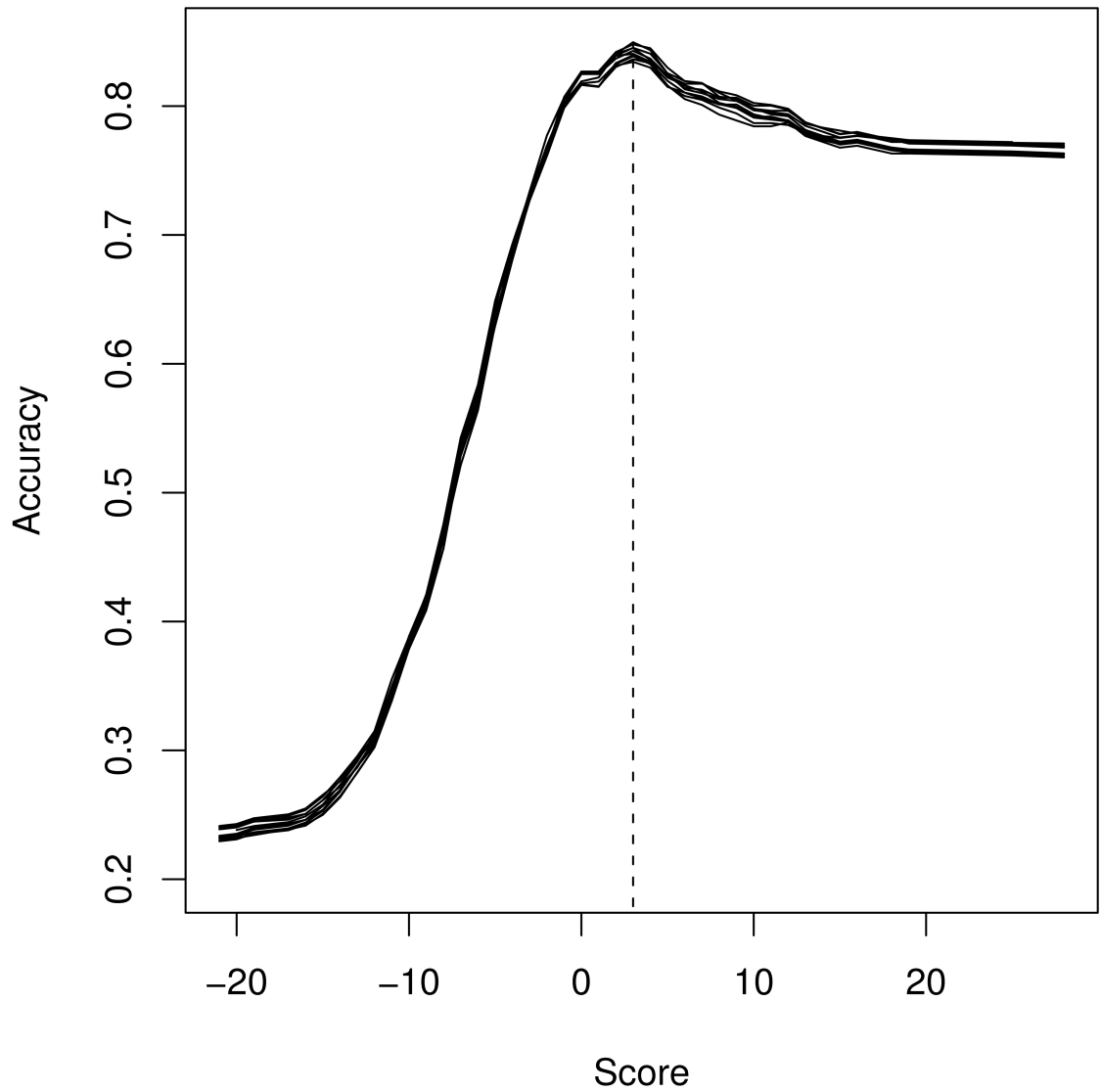

**Figure 2.** Calibration plot. The exact threshold value (θ = 3) is obtained by averaging maximal accuracy over 10 runs of cross-validation process.